\documentclass[twocolumn,aps]{revtex4}
\usepackage{amsmath}
\usepackage{amssymb}
\usepackage{lipsum}
\usepackage{graphicx}
\usepackage{dcolumn}
\usepackage{bm}
\usepackage{xcolor}
\usepackage{epstopdf}
\usepackage[utf8]{inputenc}
\DeclareUnicodeCharacter{2212}{-}

\begin{document}

\preprint{Phys.Rev.Research}

\title{Interaction-controlled transport in a two-dimensional massless-massive Dirac system:
Transition from degenerate to nondegenerate regimes}

\author{ A. D. Levin,$^1$ G. M. Gusev,$^1$ F. G. G. Hernandez,$^1$ E. B. Olshanetsky,$^{2,3}$ V. M. Kovalev,$^{2,4,5}$ M. V. Entin$^{2,3}$,  and  N. N. Mikhailov$^{2,3}$}

\affiliation{$^1$Instituto de F\'{\i}sica da Universidade de S\~ao
Paulo, 135960-170, S\~ao Paulo, SP, Brazil}
\affiliation{$^2$Institute of Semiconductor Physics, Novosibirsk
630090, Russia}
\affiliation{$^3$Novosibirsk State University, Novosibirsk 630090,
Russia}
\affiliation{$^4$Novosibirsk State Technical University, Novosibirsk 630073,
Russia}
\affiliation{$^5$Abrikosov Center for Theoretical Physics, Moscow Institute of Physics and Technology, Dolgoprudny, 141701, Russia}

\date{\today}
\begin{abstract}
The resistivity of two-dimensional (2D) metals generally exhibits insensitivity to electron-electron scattering. However, it's worth noting that Galilean invariance may not hold true in systems characterized by a spectrum containing multiple electronic branches or in scenarios involving electron-hole  plasma. In the context of our study, we focus on 2D electrons confined within a triple quantum well (TQW) based on HgTe. This system displays a  coexistence of energy bands featuring both linear and parabolic-like spectra at low energy and, therefore, lacks the Galilean invariance. This research employs a combined theoretical and experimental approach to investigate the transport properties of this two-component system across various regimes. By manipulating carrier density and temperature, we  tune our system from a fully degenerate regime, where resistance follows a temperature-dependent behavior proportional to $T^2$, to a regime where both types of electrons adhere to Boltzmann statistics. In the non-degenerate regime, electron interactions lead to resistance that is weakly dependent on temperature. Notably, our experimental observations closely align with the theoretical predictions derived in this study. This work establishes the HgTe-based TQW as a promising platform for exploring different interaction dominant scenarios for the massless-massive Dirac system.
\end{abstract}

\maketitle
\section{Introduction}
The impact of electron-electron scattering on the transport characteristics of various two dimensional (2D) conductors, which do not adhere to Galilean invariance, has garnered significant attention over the years \cite{pal}. These studies typically focus on systems that involve two distinct types of charge carriers, characterized by differing charges or effective masses. The presence of two different charge carriers within the system, each with distinct masses, poses a challenge to the traditional concept of Galilean invariance. Consequently, the direct proportionality between the net current and the total particle momentum is no longer upheld \cite{nagaev1, nagaev2, murzin, kravchenko, hwang} .
\begin{figure}[ht]
\includegraphics[width=9cm]{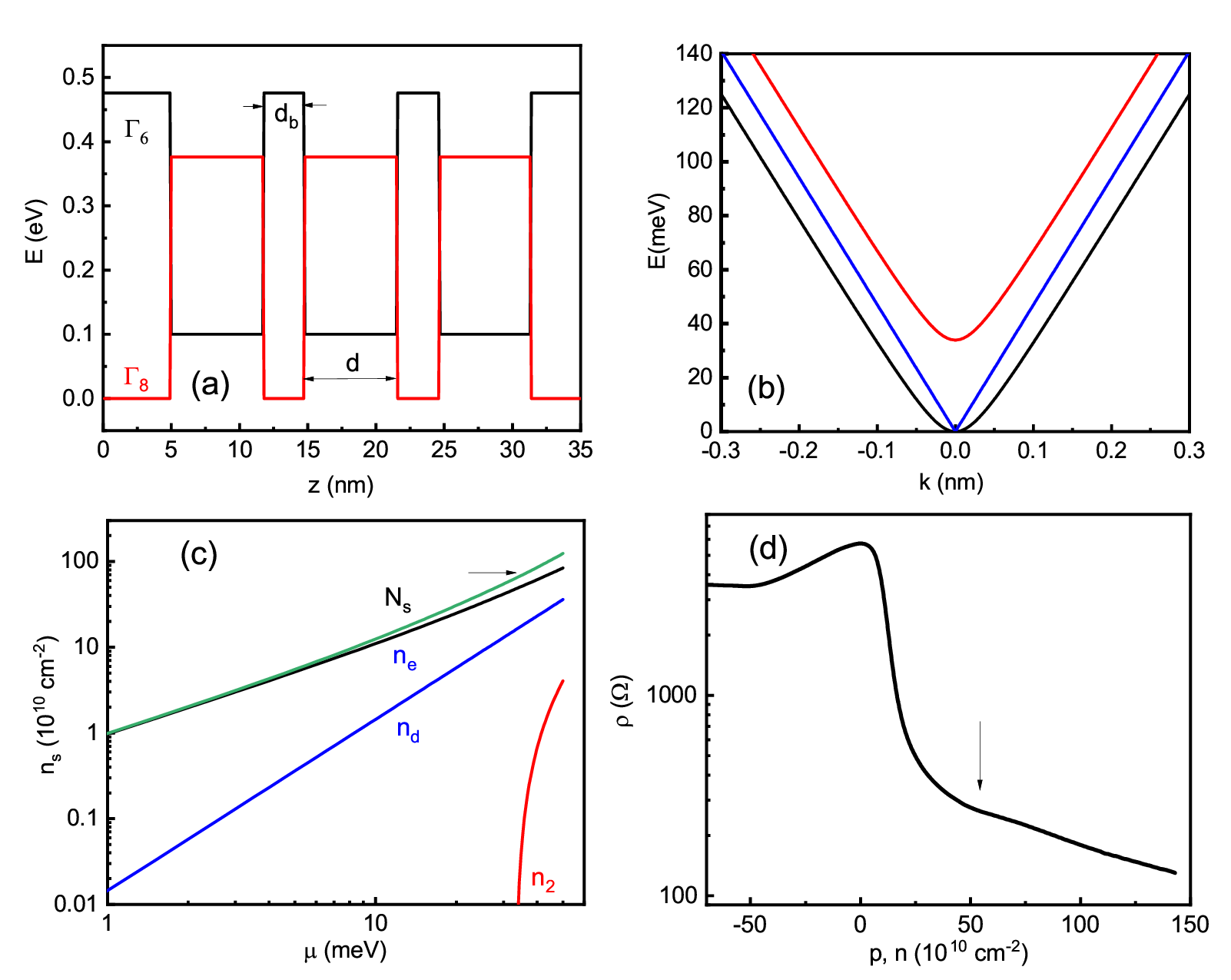}
\caption{(Color online) (a) The conduction and valence band edges of the triple quantum well are schematically shown. The widths d of the HgTe wells and the thickness $d_{b}$ of the $Hg_{1−x}Cd_{x}Te$ barriers (x=0.3) are indicated.
(b) The band structure of the triple HgTe well at $E > 0$ is calculated using the tight-binding model described in the text. The black and blue lines represent dispersion curves for massless and massive Dirac fermions.
(c) The density of carriers is shown as a function of the chemical potential. $n_{e}$ and $n_{d}$, representing the densities of massive and massless electrons, $N_{s}$ is the total density. An arrow indicates the beginning of the $E^{+}$ subband populations.
(d) The resistance of the 6.4 nm sample is plotted as a function of carrier density at T=4.2 K.}
\end{figure}
Exploring the unique scenarios of strong friction between electrons and holes, particularly in degenerate 2D semimetals, has led to the observation of resistivity proportional to $T^{2}$ \cite{olshanetsky, entin}. These phenomena have been validated in various experimental settings, most notably in HgTe quantum wells. Moreover, investigations into the nondegenerate regime have ventured into systems like single-layer and bilayer graphene, where thermally excited electron-hole pairs play a crucial role \cite{tan}. Recent advancements have also revealed electron-hole friction behavior in bilayer graphene systems, showcasing resistivity scaling as $T^{2}$  with spatially separated electrons and holes \cite{bandurin}.

In a system comprising two subbands with significantly differing masses, it can be anticipated that the resistivity will exhibit a pronounced increase with rising temperature. This increase occurs due to the relationship between the resistivity limits at high and low temperatures, which is proportional to the effective mass ratio, as described in \cite{pal}. In such systems, the dominance of interactions in the transport process surpasses the Drude resistivity resulting from impurity or phonon scattering. The behavior of transport governed by interactions at elevated temperatures is characterized by the principles of hydrodynamics and is often referred to as electronic fluid behavior.

Another crucial question pertains to the investigation of interparticle collisions and hydrodynamic conductivity within a highly adjustable system. This system can be fine-tuned, transitioning from a non-degenerate Boltzmann regime to a degenerate Fermi liquid regime.

As the understanding of electron hydrodynamics continues to unfold, it holds the promise of shedding light on the intricate interplay between particle-particle interactions, sample geometry, and charge carriers' distinct properties. This realm not only presents fundamental scientific inquiries but also offers potential applications in future electronic devices and technologies.

Previously, we offered a system that serves as a useful platform for studying transport dominated by interaction in 2D conductors \cite{gusev}. This system is a  6.3 nm HgTe quantum well, representing the spectrum as a single-valley Dirac cone near the zero-energy state. However, it's important to note that below Dirac zero point energy, the spectrum represents laterally located heavy hole valleys with a minimum at a non-zero wave vector\cite{gusev, buttner, kozlov, gusev2, kristopenko, kvon}. Consequently, when the chemical potential in the valence band reaches this lateral heavy hole band, the Dirac holes experience strong scattering by heavy holes, and such scattering violations of Galilean invariance lead to a strong $T^2$ dependence of resistance. An interesting and important aspect of physics is explored in such conductors when heavy particles are partially degenerate: Dirac holes follow Fermi statistics, while heavy holes adhere to Boltzmann statistics. It's worth noting, however, that this system does not allow for a complete study of the transition from fully degenerate to non-degenerate regimes for all subsystems \cite{gusev}.

In this current study, we present an experimental situation that enables such an investigation. We have introduced the HgTe-based triple quantum well (TQW) as a convenient system featuring two subbands with both massless and massive Dirac fermions, specifically electrons. This stands in stark contrast to the singular HgTe well explored in a prior study \cite{gusev}, where the interaction between Dirac and heavy holes resulted in significant scattering. TQW affords us the opportunity to investigate hydrodynamic conductivity across various regimes, including the strongly degenerate Fermi and non-degenerate Boltzmann regimes.  Therefore, we illustrate that regardless of the carriers' sign, spectrum type, and other characteristics, including the confinement features of the multilayer system, electron-electron collisions exhibit unified properties and can emerge as the dominant mechanism in ultraclean systems. Consequently, TQWs emerge as a promising platform for exploring novel phenomena resulting from interactions. Notable instances encompass the violation of the Wiedemann-Franz law, as reported in paper\cite{lucas}, the prediction of giant magnetoresistance in work \cite{levchenko}, and quantum critical conductivity, detailed in ref. \cite{muller}, among other significant theoretical predictions.

\section{ELECTRON SPECTRUM  IN A TRIPLE-WELL SYSTEM}
HgTe-based quantum wells have garnered significant interest due to their capability to create unconventional two-dimensional (2D) systems, such as 2D topological insulators \cite{konig, hasan, kvon2, gusev6}. Additionally, the spectrum's behavior is primarily determined by the well's thickness, leading to various phases characterized by insulating gaps, gapless regions, and inverted subbands \cite{gerchikov,kane, bernevig,bernevig2}. 
Double and triple quantum wells are multi-layer systems comprising two or three quantum wells separated by a tunneling-transparent barrier. Theoretical  investigations \cite{kristopenko2, michetti, ferreira} propose that the phase states within these structures are significantly altered when compared to the single quantum well scenario, which is supported by experimental evidence \cite{gusev3, gusev4}. This alteration leads to a more complex phase landscape due to the additional degrees of freedom introduced by the increased number of 2D subbands and the hybridization induced by tunneling between them. Here, we illustrate that a triple HgTe-based quantum well serves as a valuable platform for studying the influence of dominant interactions on transport behavior, primarily due to the unique characteristics of its spectrum.

Figure 1a depicts the band structure of triple HgTe quantum wells with a thickness of $d_{b} = 3 nm$ and a well width of $d = 6.7 nm$. The schematic representation displays the conduction and valence band edges of the triple quantum well.

In  paper  \cite{ferreira}, the authors consider the confinement characteristics of the subbands within triple HgTe/CdTe quantum wells, along with their corresponding topological properties and edge state attributes. To accomplish this, they employ an effective 2D Hamiltonian for the HgTe-based triple well system.

For single HgTe quantum wells, this is achieved by projecting the Hamiltonian onto its eigenstates at $k_{II} = 0$, leading to the well-established BHZ model \cite{bernevig, bernevig2}. In contrast, for double HgTe quantum wells, two intriguing approaches are explored. Firstly, similar to the derivation of the BHZ Hamiltonian, the authors of the paper referenced as \cite{kristopenko2} project the overall Hamiltonian onto the eigenstates at $k_{II} = 0$ for the double quantum wells (DQW).

Alternatively, in Reference \cite{michetti}, the authors project the total DQW Hamiltonian onto the subbands of the individual wells (left and right) and introduce tunneling parameters to account for the coupling between neighboring quantum wells.

In  paper \cite{ferreira} the authors investigate the phase diagram of the HgTe triple well and transform the 3D Kane Hamiltonian into an effective 2D 3×BHZ model. This transformation enables them to explore the edge state characteristics in each topological phase.

These findings reveal the existence of gapless phases, attributed to the slight hybridization of H-like states from different quantum wells. In these gapless phases, one or two pairs of edge states are present within the bulk. However, there is also a phase where all E-like and H-like subbands are inverted, leading to the formation of three sets of edge states within a bulk gap.

Moreover, for a triple well with a well width of $d = 6.7 nm$, the subbands $E_{01}$ and $H_{01}$ exhibit significant hybridization resulting in nearly negligible mass for a linear Dirac subband. Meanwhile, the second subband maintains a k-parabolic spectrum with massive Dirac fermions at low energies. This effect has been substantiated through the measurement of Schubnikov de Haas oscillations \cite{ferreira}.

To obtain the analytical expression for the triple well spectrum, we employ a tight-binding model with parameters calibrated based on Kane model calculations, as detailed in \cite{ferreira}. In this case the effective low energy Hamiltonian is
\begin{equation}\label{eq1}
  H = 
  \begin{pmatrix}
    0 & Ak & 0 & 0& 0 & 0\\
    Ak & 0 & t & 0& 0 & 0\\
    0 & t & 0& Ak& 0 & t\\
    0 & 0 & Ak & 0& 0 & 0\\
    0 & 0 & 0 & 0& 0 & Ak\\
    0 & 0 & t & 0& Ak & 0 \\
  \end{pmatrix}
\end {equation}
where $k$ is the two-dimensional
momentum operator, $A=\hbar v_{F}\approx 470$ meV nm,  $v_{F}$ is the Fermi velocity of the single HgTe well and the in-plane nearest-neighbor hopping is $t \approx 24 meV$.

One can see that the Hamiltonian, (1), leads to a combination of two linear  bands  and four massive bands:

\begin{equation}\label{eq2}
\begin{gathered}
E_s= \pm\left[t^2+v_F^2 k^2+s \sqrt{t^4+2 t^2 v_F^2 k^2}\right]^{1 / 2}, \quad s= \pm 1 \\
E_0= \pm v_F k .
\end{gathered}
\end {equation}
In Figure 1b, we present the electronic segment of the spectrum. The Dirac-like spectrum is represented by the blue lines, whereas the massive Dirac subbands are denoted by the black and red lines. It's worth noting that the simplified tight-binding model provides a good fit for the electronic branches, consistent with those obtained using the Kane model \cite{ferreira}. However, there is a notable difference for the hole segments of the spectrum. Therefore, in our work, we concentrate solely on the electronic spectrum.

It's noteworthy that in symmetric HgTe TQWs, the transition from a gapless to a gapped phase can be achieved by disrupting the inversion symmetry of the wells through the application of a transverse electric field, either from asymmetric doping or external gate bias \cite{michetti, gusev3}. This process can induce a small gap in the energy spectrum, reminiscent of phenomena seen in bilayer and trilayer graphene. The ability to manipulate this gap with an external field is fascinating, offering avenues to modulate edge state transport within HgTe double and triple well configurations. As we will show in the next paragraph, we detected a minimal gap in our structures, indicated by activation resistance behavior, in the order of approximately 1 meV. However, it's critical to underline that this gap, while present, is minuscule relative to other system parameters and does not notably affect the spectrum as outlined by Equation 1. Our investigation primarily concentrated on the conduction band, which is essentially unaffected by such a negligible gap.

Due to the Dirac-like nature of the Hamiltonian, the behavior of the density of states significantly deviates from that of conventional 2D systems. In the spectrum of subbands denoted by $E_s$, we observe a nearly parabolic shape, with particles exhibiting significant mass only in the vicinity of $k \approx 0$. As energy levels increase, the spectrum also transitions to a linear regime. This leads to a linear dependence of the density of states on energy and a non-monotonic change in the chemical potential with respect to gate voltage.

In Figure 1c, the density of 2D carriers, distributed across different subbands, is depicted as a function of the chemical potential $\mu$. Notably, it is evident that the density of carriers in the massive Dirac branch surpasses that of the massless carriers. Furthermore, beyond an energy threshold of 35 meV, the second subband of the massive Dirac fermions starts to become occupied.

The presence of various electronic branches leads to the breakdown of Galilean invariance and a noteworthy increase in the impact of scattering between these branches on electronic transport.

\begin{figure*} 
  \centering
\includegraphics[width=15cm]{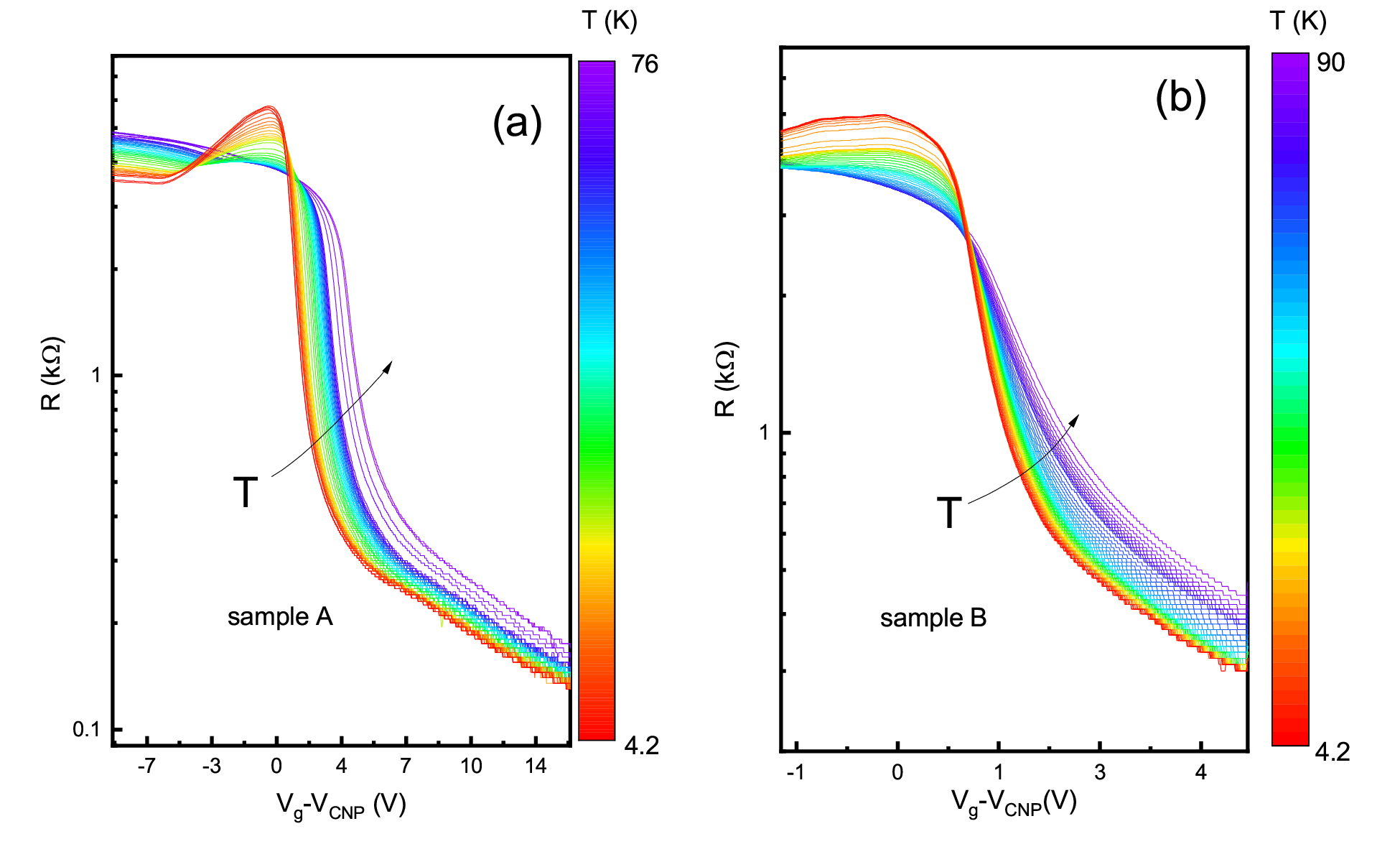}
\caption{(Color online) Resistance as a function of the gate voltage at
different temperatures for two HgTe triple quantum wells. }
\end{figure*}
\section{Experimental results}
We fabricated triple quantum wells using $HgTe/Cd_{x}Hg_{1-x}$Te material with a [013] surface orientation. The wells had equal widths, with $d_0$ measuring 6.7 nm and a barrier thickness of $t$ set at 3 nm. The layer thickness was monitored during MBE growth via ellipsometry, achieving an accuracy within ± 0.3 nm.
\begin{figure*} 
  \centering
\includegraphics[width=15cm]{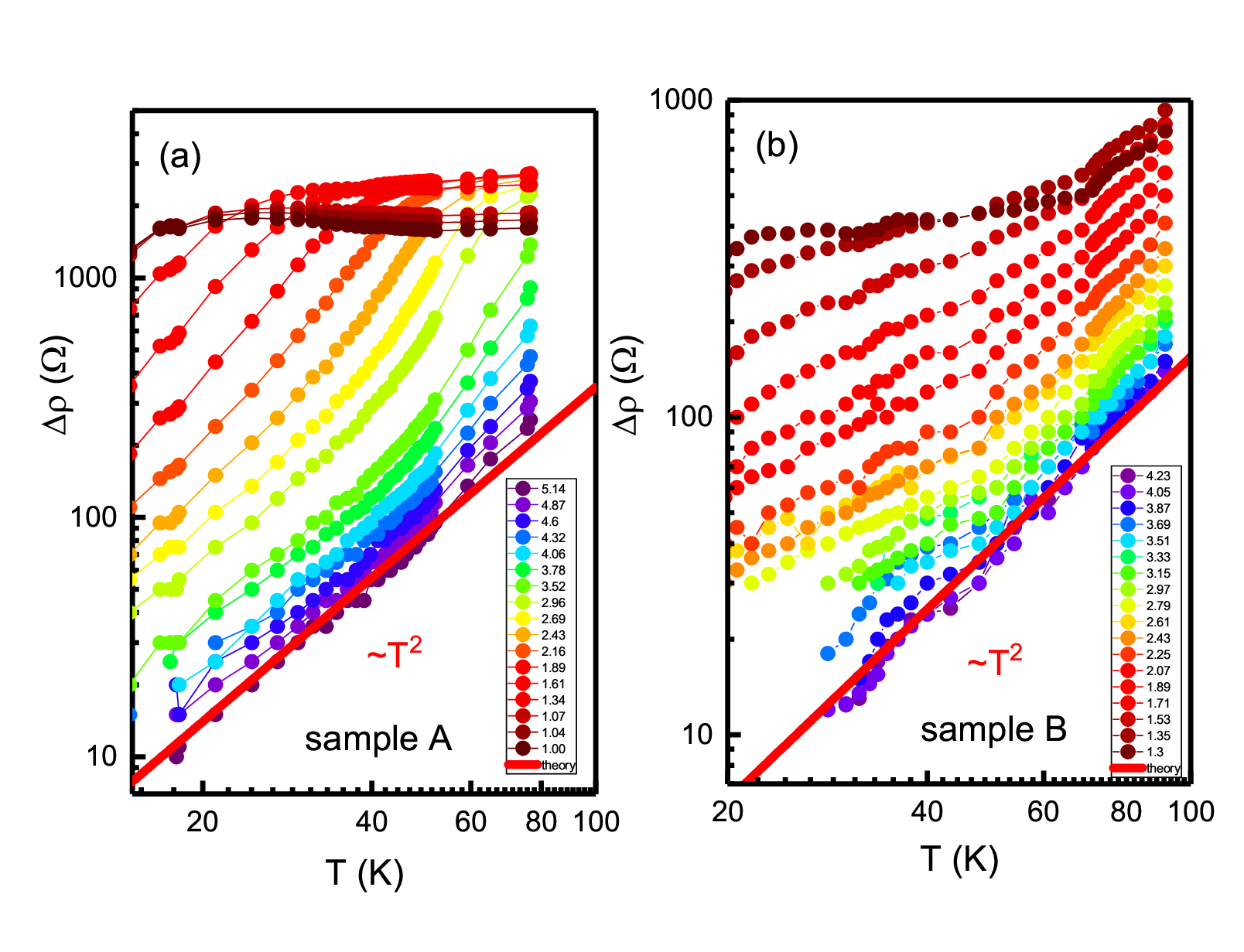}
\caption{(Color online) Excess resistivity $\Delta
\rho(T)=\rho(T)-\rho(T=4.2K)$ as a function of the temperature for various densities for sample A (a) and B (b). The red lines show $T^{2}$ dependence. The values of the densities is given in $10^{11} cm^{-2}$.}
\end{figure*}
\begin{figure*} 
  \centering
\includegraphics[width=15cm]{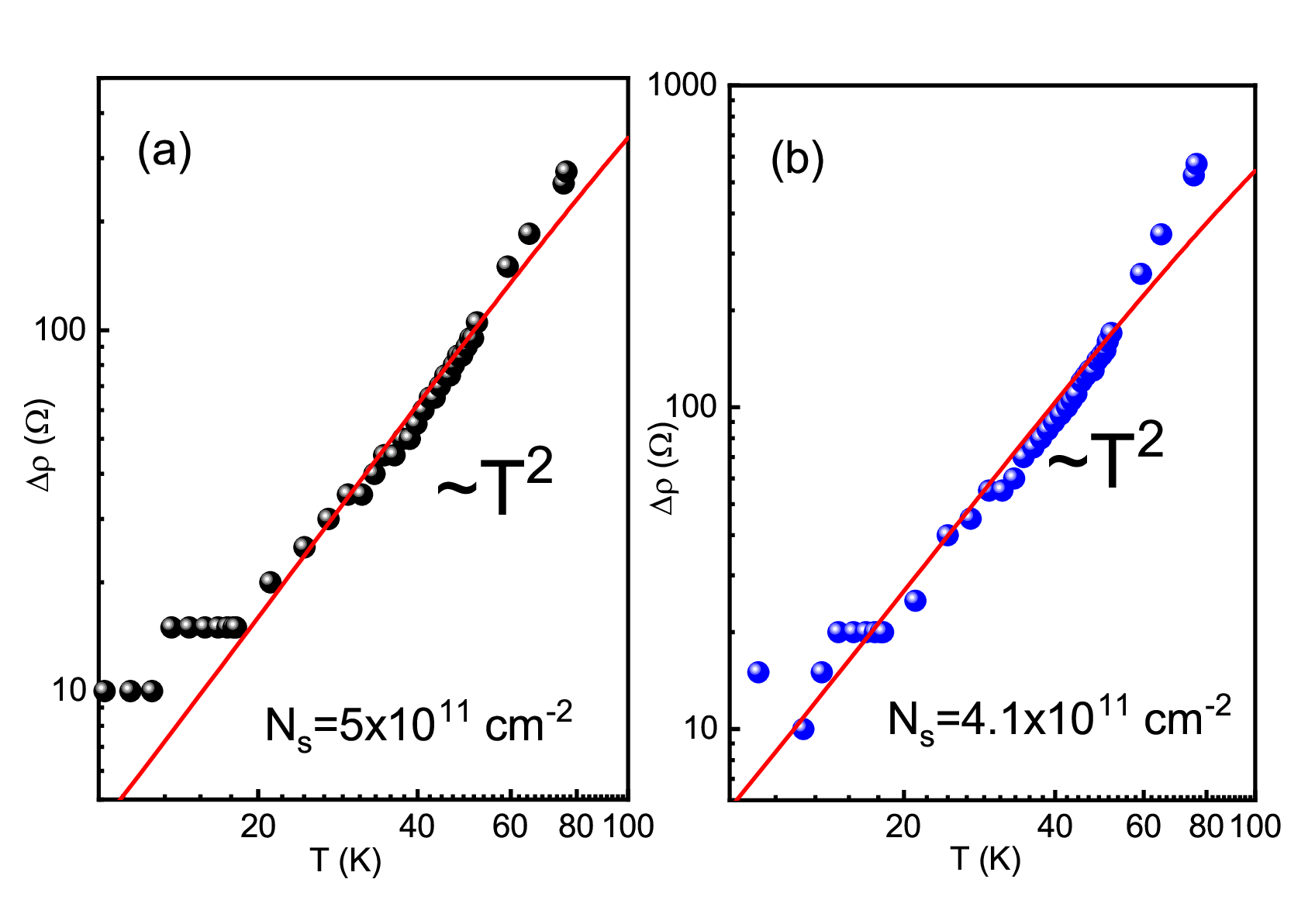}
\caption{(Color online) Excess resistivity $\Delta
\rho(T)=\rho(T)-\rho(T=4.2K)$ as a function of the temperature for two different densities for sample A. (a) The total density is
$N_{s}=5\times10^{11} cm^{-2}$; (b) $4.1\times10^{11} cm^{-2}$. Circles represent the experimental data, and the red lines represent the theory of interparticle scattering between the fully degenerate massless and massive electrons. }
\end{figure*}
The devices employed in this study were multiterminal bars featuring three consecutive segments, each 3.2 $\mu m$ wide, with varying lengths of $2 \mu m$, 8 $\mu m$, and $32 \mu m$. These devices were equipped with nine contacts. The contacts were created by indium bonding to the surface of the contact pads, which were precisely defined using lithography. Given the relatively low growth temperature (around 180°C), the temperature during the contact fabrication process remained low as well. Indium diffused vertically downward on each contact pad, establishing an ohmic connection across all three quantum wells, with contact resistance falling within the 10–50 $k \Omega$ range.
\begin{figure*} 
  \centering
\includegraphics[width=15cm]{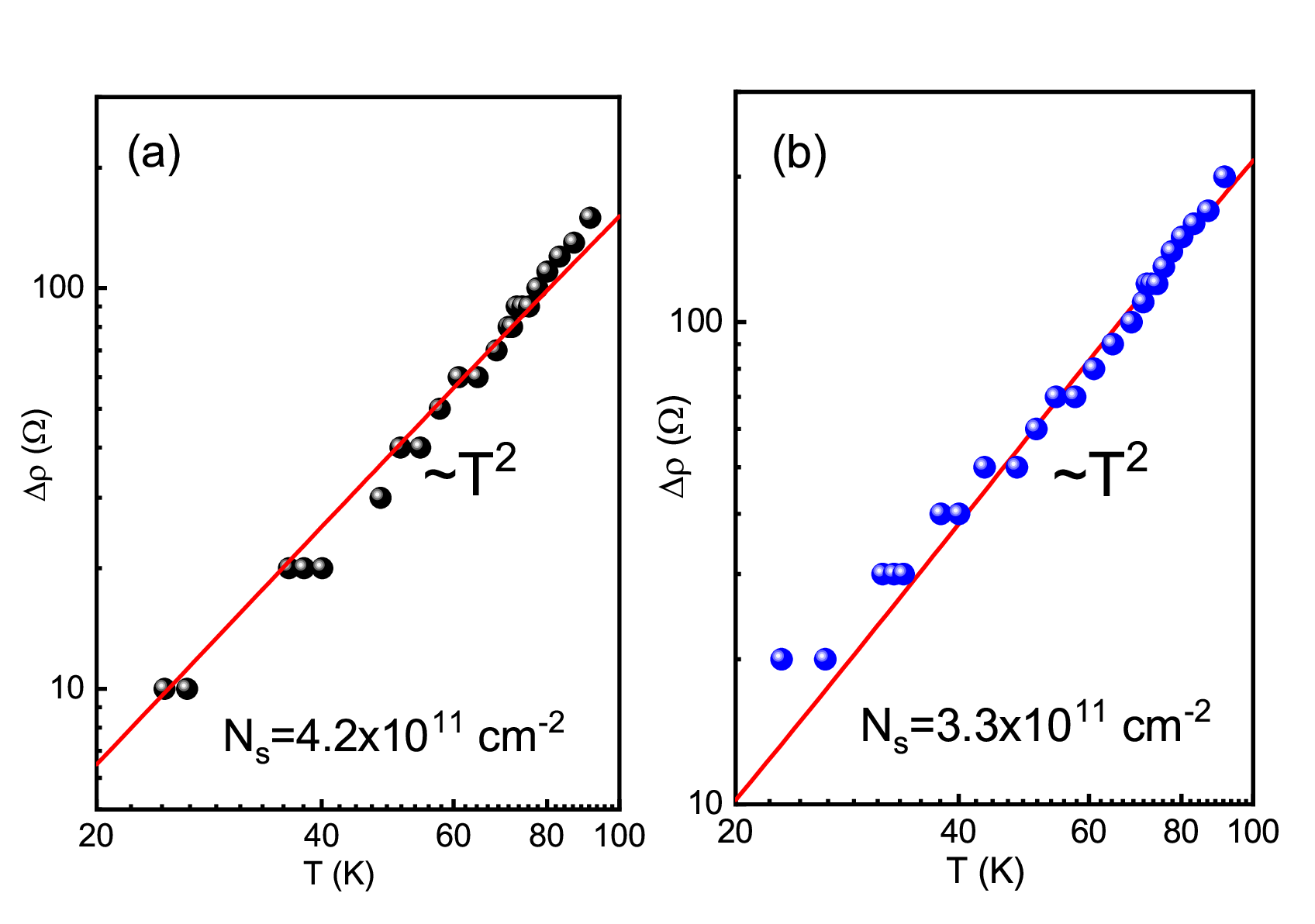}
\caption{(Color online) Excess resistivity $\Delta
\rho(T)=\rho(T)-\rho(T=4.2K)$ as a function of the temperature for two different densities for sample B. (a) The total density is
$N_{s}=4.2\times10^{11} cm^{-2}$; (b) $3.3\times10^{11} cm^{-2}$. Circles represents the experimental data, the  the red lines represent the theory of interparticle scattering between the fully degenerate massless and massive electrons. }
\end{figure*}
Throughout the AC measurements, we consistently verified that the reactive component of impedance did not exceed $5\%$ of the total impedance, confirming the effectiveness of the ohmic contacts. Furthermore, the current-voltage (I-V) characteristics exhibited ohmic behavior at low voltages. A 200 nm $SiO_{2}$ dielectric layer was deposited onto the sample surface, subsequently covered by a TiAu gate. The density variation with gate voltage was estimated to be approximately $ 0.9 \times 10^{11} cm^{-2}/V$, derived from the dielectric thickness and Hall measurements, as previously reported in studies employing similar devices. Two samples, denoted as A and B  were studied.
The samples were created from a uniform substrate and were subject to identical growth conditions. It is important to emphasize that these samples are of mesoscopic dimensions and, as such, are expected to display all characteristics inherent to mesoscopic physics.
\begin{table}[ht]
\centering
\begin{tabular}{|l|l|l|l||l||l||l|}
\hline
sample & $d$ (nm) & $V_{CNP}$ (V)& $\rho_{max}(h/e^{2})$& $\mu_{e} (V/cm^{2}s )$ \\
\hline
A&    6.7 & -3.6& 0.22 & 33.600\\
\hline
B&   6.7 & -6.7 & 0.2 &39.600\\
\hline
\end{tabular}
\caption{\label{tab1} Some of the typical parameters of the electron system in HgTe triple quantum well at T=4.2K.}
\end{table}

Figure 1d illustrates the variation of resistivity with charge density for sample A at a temperature of 4.2 K. The resistance shows a prominent peak centered at the charge neutrality point, which corresponds to the zero-energy point. This behavior is a characteristic feature of both gapless and gapped HgTe single well and double well devices \cite{kvon, gusev3}. The maximum electron density corresponds to a Fermi energy value of approximately 150 meV, as depicted in Figure 1c. As the chemical potential approaches the range of 30-40 meV, the second subband associated with massive Dirac fermions begins to be occupied. This is evident from a small feature in the resistivity, as indicated by the arrow in Figure 1d. 
Table 1 presents the key parameters of the gapless HgT quantum well employed in this research. These parameters include the well width (d), the gate voltage associated with the Dirac point position ($V_{CNP}$), the resistivity ($\rho$ value) at the charge neutrality point (CNP), and the electron mobility ($\mu_{e}$) calculated as $1/\rho N_{s}$, where the total electron density ($N_{s}$) is set at $2\times10^{11}$ cm$^{-2}$.

Figure 2 illustrates the variation in resistance with respect to gate voltage across a wide range of temperatures. The plot reveals a notable increase in resistance as temperature rises, with one notable exception: in the voltage region close to the charge neutrality point ($-3 V < V_g - V_{CNP} < 0 V$), the resistance exhibits insulating behavior characterized by a very slight activation gap in the range of 0.5-0.8 meV. The reduction in resistance as temperature increases can be attributed to potential fluctuations, which may lead to the creation of small semimetallic insulating regions or residual gaps. It is important to acknowledge that even a small gap has the potential to alter the energy spectrum described by Equation (2). However, for the sake of simplicity, we have omitted such modifications in our straightforward model.

To further investigate the temperature-dependent behavior of resistance (or resistivity), we calculate the excess resistivity, denoted as $\Delta\rho(T) = \rho(T) - \rho(T=4.2K)$. 
Subsequently,  in Figure 3, we present the excess resistivity for different electron densities across a broad temperature range, for samples A and B. It is evident that the temperature dependence evolves, adhering closely to a $T^{2}$ relationship at high electron densities, while exhibiting weak temperature dependence near the charge neutrality point. It is noteworthy that in the intermediate density regime, the temperature dependence displays a somewhat more complex behavior. At low temperatures, the dependence in sample A resembles a quadratic function, whereas at higher temperatures, the power becomes significantly more pronounced. In contrast, for sample B, the temperature dependence remains quadratic at high temperatures, but the power is diminished at lower temperatures.  At low densities, sample A exhibits an almost plateau-like behavior at high temperatures, while sample B demonstrates an increase in excess resistivity. We believe that sample B displays greater levels of disorder and inhomogeneity, and that contributions from regions of higher density become critical at elevated temperatures. At high densities, the experimental curve follows a quadratic dependence with a difference in the order of amplitude for both samples. In the following discussion, we focus on the high-density dependencies for a comparison with theoretical predictions.        
Figure 4 an d 5 depicts the temperature dependence of $\Delta\rho(T)$ across two densities for sample A and sample B respectively in more details. It is evident that $\rho(T)$ exhibits a $T^2$ dependence across all depicted densities for both samples, with some minor deviations. Notably, we observed deviations in sample B at lower densities, which we attributed to density inhomogeneity. It is reasonable, because we observed that the charge neutrality point in sample B shifted towards a higher negative voltage, likely due to charge capture by dielectrics. Furthemore, the experimental data points at low temperatures exhibit dispersion due to their limited precision.  In sample B, we also observed a more rapid growth of excess resistivity with increasing temperature, which deviated from the $T^2$ trend, possibly owing to the broadening of the distribution function. We will delve further into the Boltzmann statistical regime in the subsequent section.

The distinct resistance dependence of $T^{2}$  serves as an unambiguous  indicator of electron-electron scattering, as opposed to phonon scattering, which would lead to a linear, rather than quadratic, temperature dependence, as discussed in work \cite{melezhik}. While it is a well-established fact that electron-electron interactions do not influence the resistivity of a Galilean-invariant Fermi liquid, there are specific scenarios where a conductivity proportional to $T^{-2}$ can be observed, as detailed in review \cite{pal}. Such scenarios include  the presence of spin-orbit interactions or the involvement of multiple subbands, as discussed in the studies  \cite{nagaev1, appel, murzin, kravchenko, nagaev2, pal}.

Notably, Figure 2 reveals an extraordinary observation: the resistivity ratio between high and low temperatures, $\rho(T=70K)/\rho(T=4.2K)$, surpasses the 4-5 limit, signifying that $\Delta\rho$ is significantly greater than $\rho(T=4.2K)$ at elevated temperatures. This striking departure from more conventional scenarios, where resistance is predominantly influenced by disorder or phonon scattering, underscores the critical role of particle-particle collisions in this context. These collisions have a pronounced impact, far exceeding that of impurity-related scattering. This observation validates the HgTe-based triple well as a promising experimental platform for investigating transport phenomena dominated by interactions, opening the door to the exploration of novel and non-trivial effects, such as violations of the Wiedemann-Franz law \cite{sarma}  anomalous Coulomb drag \cite{liu} and many others (for review, see \cite{polini}).

\section{Comparison of theory with experiment}
In the following discussion, we examine a straightforward hydrodynamic model with two subbands, wherein electron-electron scattering is characterized by a concept of mutual friction. The conductivity is determined through an equation of motion for the electron featuring Dirac dispersion and massive electrons.
\begin {equation}\label{eq3}
-\frac{v_d-v_e} {\tau_{de}}  - \frac {v_d} {\tau_d} + \frac{q E}{m_d}= 0;
\end {equation}
\begin {equation}\label{eq4}
-\frac{v_e-v_d} {\tau_{ed}}  -\frac{v_e}{\tau_e} + \frac{q E}{m_e}=0;
\end {equation}
where variables $v_d$ and $v_e$ are drift velocities of the massless (Dirac) and massive electrons. 
The solution of this equations has been performed elsewhere \cite{gusev,entin, bandurin}. 
The expression for conductivity can be obtained with some modifications, and it is given by:
\begin{widetext}
\begin{equation}\label{eq5}
\sigma=q^2 \frac{n_d n_e\left[  \tau_d  \left(\tau_e+\tau _{int}\right)\left(2+ \frac{n_d}{n_e}\right)+
\tau _e \left(\tau_d+\tau _{int}\right)\frac{n_e}{n_d}+\tau _{int}\left(\frac{m_e}{m_d} \tau_d-\frac{m_d}{m_e}\tau _e\right)\right]}
{m_e n_e \left(\tau _d+\tau _{int}\right)+m_d n_d \left(\tau _e+\tau _{\text{int}}\right)}
\end{equation}
\end{widetext}
We introduce the following variables: $n_{e}$ and $n_{d}$, representing the densities of massive and massless electrons, respectively, and $\tau_{e}$ and $\tau_{d}$ as their respective scattering times, due to impurities and static defects. Furthermore, we define $N_{s}$ as the total electron density. In addition, $q$ is the elementary charge, $m_{e}=\hbar^2 k \frac{dk}{dE_{s}}$ is the heavy electron effective mass, $m_{d}=\mu/v_{F}^{2}$ is the Dirac electron effective mass, $\tau_{ed (de)}$ is the collision time with heavy (Dirac) electron per Dirac (heavy) electron. We also introduce $1/\tau_{int}=1/\tau_{de}+1/\tau_{ed}$. 
The equation, considering the constraint that interactions between massive and massless Dirac electrons conserve the overall momentum density, is: $n_d m_d \tau_{d e}=n_e m_e\tau_{e d}$.
Hence, the term in equation 5 that remains temperature-independent is anticipated to be primarily governed by interface roughness and impurity scattering, as discussed in  work \cite{melezhik}. Conversely, the temperature-dependent component of resistivity can be attributed to electron-electron (e-e) friction within the non-Galilean invariant massive-massless Dirac liquid.

An insightful analysis would involve examining the limits of $T=0$ and $T=\infty$ especially in cases where the effective masses differ significantly. The Dirac hole's effective scattering time can be approximated using the relation $m_{d}=\mu/v_{F}^{2}\approx0.006 m_{0}$, with $\mu\approx 16 meV$.
At $T\rightarrow 0$ both bands contribute to the total conductivity dominated by the Dirac holes:
\begin{equation}\label{eq6}
\sigma(T=0)=\frac{q^{2}n_{d}v^{2}\tau_{d}}{\mu}=\frac{q^{2}n_{d}\tau_{d}}{m_{d}}.
\end{equation}

At $T=\infty$ the conductivity becomes temperature independent and saturates at a value approximately determined by the conductivity of the heavy hole band :
\begin{equation}\label{eq7}
\sigma(T=\infty)=\frac{q^{2}(n_{d}+n_{e})^{2}\tau_{e}}{m_{e}n_{e}}
\end{equation}
In our case $n_{e}>>n_{d}$, the ratio of the resistivities at both temperature limits is determined by :
\begin{equation}\label{eq8}
\frac{\rho(T=\infty)}{\rho(T=0)}=\frac{m _{e}n_{d}\tau_{d}}{m_{d}n_{e}\tau_{e}}
\end{equation}
When  $n_{e}\tau_{e} \sim n_{d}\tau_{d}$, the ratio of the resistivities at these temperature extremes is predominantly influenced by the effective mass ratio (for Fermi energy $E_{F}=5 meV$) \cite{pal}:

\begin{equation}\label{eq9}
\frac{\rho(T=\infty)}{\rho(T=0)}\approx\frac{m _{e}}{m_{d}}\approx4.4
\end{equation}
Hence, it is reasonable to anticipate that in the optimal system where electron-electron interactions govern transport, there will be energy branches characterized by distinct effective masses. The gapless HgTe system stands out as a highly promising platform for conducting this type of research.

In a regime of full degeneracy, where both types of electrons adhere to Fermi statistics, an expression for particle-particle collisions has been derived, with slight modifications to account for the difference in their energy spectra \cite{gusev,kovalev}:

\begin{equation}\label{eq10}
\frac{1}{\tau_{d e}}=\alpha \frac{m_e (kT)^2 U_0^2}{3 \pi \hbar^5 v v_F} \equiv \frac{1}{\tau_0}\sim T^{2}, \\
\end{equation}
\begin{equation}\label{eq11}
\quad \frac{1}{\tau_{e d}}=\frac{1}{\tau_0} \frac{\mu}{m_e v_{F}^2} \frac{n_d}{n_e}\sim T^{2}, \\
\end{equation}

where $U_0=\frac{2 \pi e^2}{\epsilon q_s}$, $\quad q_s=\frac{m e^2}{\varepsilon \hbar^2}$, $\epsilon$ is the dielectric constant of the material, v- is Fermi velocity of the massive particle, prefactor $\alpha$ represents a numerical coefficient that varies based on the specifics of the Coulomb interactions.
Note that $1/\tau_{e d} << 1/\tau_{d e}$, and conventional $T^2$ behavior for $1/\tau_{int}\sim 1/\tau_{d e}$, characteristic of a particle with parabolic dispersion, becomes evident.

Remember that, in the case of finite temperature, the electron density can be found from the equation:
\begin{equation}\label{eq12}
n_{d,e}=\int\limits_{0}^{\infty} D_{\varepsilon}^{d,e}(\varepsilon)(1+e^{(\varepsilon-\mu)/kT})^{-1}d\varepsilon
\end{equation}
where  $D_{\varepsilon}^{d,e}$ is the density of the states of the Dirac and massive electrons respectively. It is essential to emphasize that even though the density in each subband strongly relies on the temperature at high T or in the nondegenerate regime, the total density is constrained by the gate voltage, expressed as $n_{e}+n_{d}=N_{s}\sim V_{g}$. As a result, the chemical potential can be parametrically determined.

\begin{table}[ht]
\caption{\label{tab2} Fitting parameters in equations (5), (10) for 2 samples.}
\begin{ruledtabular}
\begin{tabular}{lcccccccccc}
&Sample& $E_{F}$&$N_{s}$&$\tau_{e}$ & $\tau_{d}$ & $n_{e}$  & $n_{d}$ & $\alpha$   \\
& & &$10^{11}$ &  $10^{-13} $ & $10^{-13} $ &  $10^{11}$  &  $10^{11}$      \\
\hline
& & meV&$cm^{-2}$& $s$  &  $s$ &  $ cm^{-2}$ & $ cm^{-2}$ &       \\
\hline
&A& 33.8&5 &$0.31$  & $4.7$ &  $3.3$ & $1.7$ &  0.5 \\
&A&30.4 &4.2 &$0.3$  & $4.6$ &  $2.8$ & $1.3$ &  0.7\\
&B&30.7 &4.2 &$0.29$  & $4.3$ &  $2.8$ & $1.3$ &  0.17 \\
&B&26.5 &3.3 &$0.41$  & $5$ &  $2.3$ & $1$ &  0.2  \\
\end{tabular}
\end{ruledtabular}
\end{table}
To compare with the theoretical framework, we performed a fitting analysis on the temperature-dependent data, as illustrated in Figures 4 and 5. This analysis involved a single adjustable parameter denoted as $\alpha$, which accounts for the interaction strength between the Dirac and massive holes as outlined in Equations \ref{eq5} -\ref{eq11}. The scattering parameters, referred to as $\tau_{e(d)}$, play a predominant role in determining resistivity at lower temperatures. Importantly, varying these parameters within a reasonable range does not impact the friction coefficient, which is the primary driver of temperature-dependent resistivity. In Figures 4 and 5 we present the theoretical dependencies of resistivity excess for various total density values. Notably, the experimental data closely aligns with the expected dependence $\Delta\rho(T)\sim T^{2}$ for the parameters specified in Table II.

It is evident that the prefactor $\alpha$, introduced as an adjustable parameter reflecting the difference between the actual Coulomb potential and the contact interaction potential, is close to unity and exhibits a tendency to increase as the particle density decreases. It's worth noting, however, that the corresponding prefactor for sample B is smaller than that of sample A. 
It is important to note that both samples exhibit a resistivity pattern consistent with the $T^2$ law within one order of magnitude.

\section{Comparison of the theory with experiment in non -degenerate regime}
\begin{figure} 
\includegraphics[width=9cm]{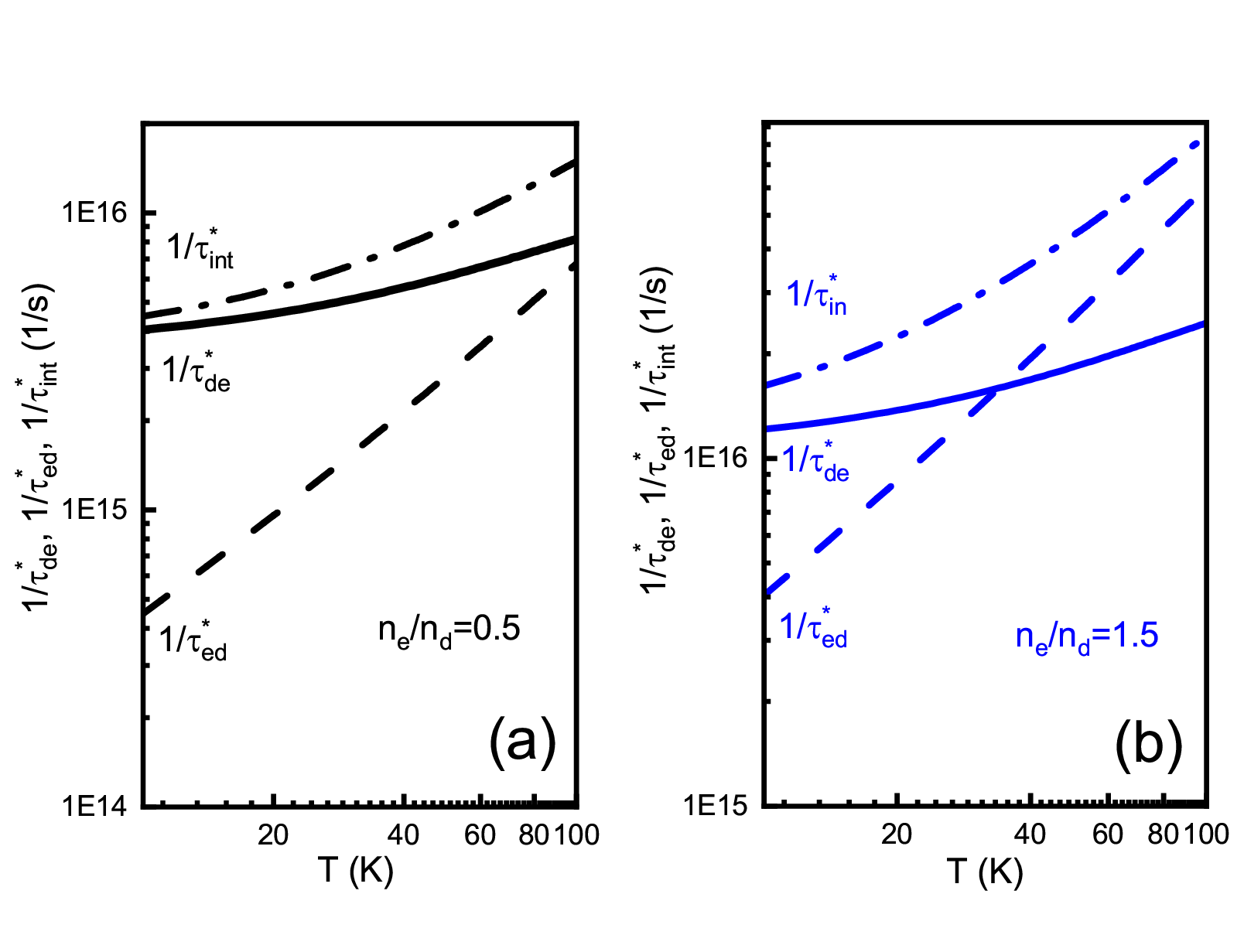}
\caption{(Color online) Temperature dependence of the relaxation rate for different density ratio $n_{e}/n_{d}$. Solid lines are computed using equations (13,15) ($1/\tau_{d e}^{*}$, dashes are computed using equations (14,16) ($1/\tau_{e d}^{*})$), dot-dashes represent $1/\tau_{int}^{*}$. Total density $N_{s}=3\times10^{10} cm^{-2}$,  (a) ratio $n_{e}/n_{d}=0.5$ (black lines), (b) $n_{e}/n_{d}=1.5$ (blue lines). }.
\end{figure}
At low densities and moderately high temperatures, both massive and massless Dirac electrons start adhering to Boltzmann statistics. In this particular regime, an expression for electron-electron collisions has been formulated, incorporating minor adjustments to accommodate variations in their energy spectra \cite{kovalev}:
\begin{equation}\label{eq13}
\frac{1}{\tau_{d e}^{*}}=\beta \frac{\sqrt{\pi}}{6}\frac{n_{e}}{(kT)^3}\frac{v_{T}m_{e}^4 v_{F}^{5}}{\hbar^{3}}U_{0}^{2}\mathfrak{J_{1}}\sim T^{-1/2}\mathfrak{J_{1}}(T)
\end{equation}
\begin{equation}\label{eq14}
\frac{1}{\tau_{e d}^{*}}=\beta \frac{\sqrt{\pi}}{4}\frac{n_{d}}{(kT)^{2}}\frac{v_{T}m_{e}^{3}v_{F}^{3}}{\hbar^{3}}U_{0}^{2}\mathfrak{J_{2}}\sim T^{1/2}\mathfrak{J_{2}}(T)
\end{equation}
where $U_0=\frac{2 \pi e^2}{\epsilon q_T}$, $\quad q_T=\frac{2 \pi e^2 n_{e}}{\varepsilon (kT)}$, $\epsilon$ is the dielectric constant of the material, $v_{F}$- is Fermi velocity of massless Dirac particles,  $v_{T}=\sqrt{\frac{2kT}{m_{e}}}$ is velocity of massive particles,  The prefactor denoted by $\beta$ corresponds to a numerical coefficient that  depends on the details of the Coulomb interactions.
Integrals $\mathfrak{J_{1,2}}$  are represented by the following expressions:
\begin{equation}\label{eq15}
\mathfrak{J_{1}}=\int\limits_{0}^{\infty}y^{2}dy\int\limits_{-1}^{1}\sqrt{1-x^{2}}e^{-\frac{m_{e}v_{F}^{2}}{2kT} (x+\frac{y}{2})^{2}}dx
\end{equation}
\begin{equation}\label{eq16}
\mathfrak{J_{2}}=\int\limits_{0}^{\infty}y^{2}dy\int\limits_{-1}^{1}\frac{e^{-\frac{m_{e}v_{F}^{2}}{2kT} (x+\frac{y}{2})^{2}}}{\sqrt{1-x^{2}}}dx
\end{equation}
\begin{figure*} 
  \centering
\includegraphics[width=15cm]{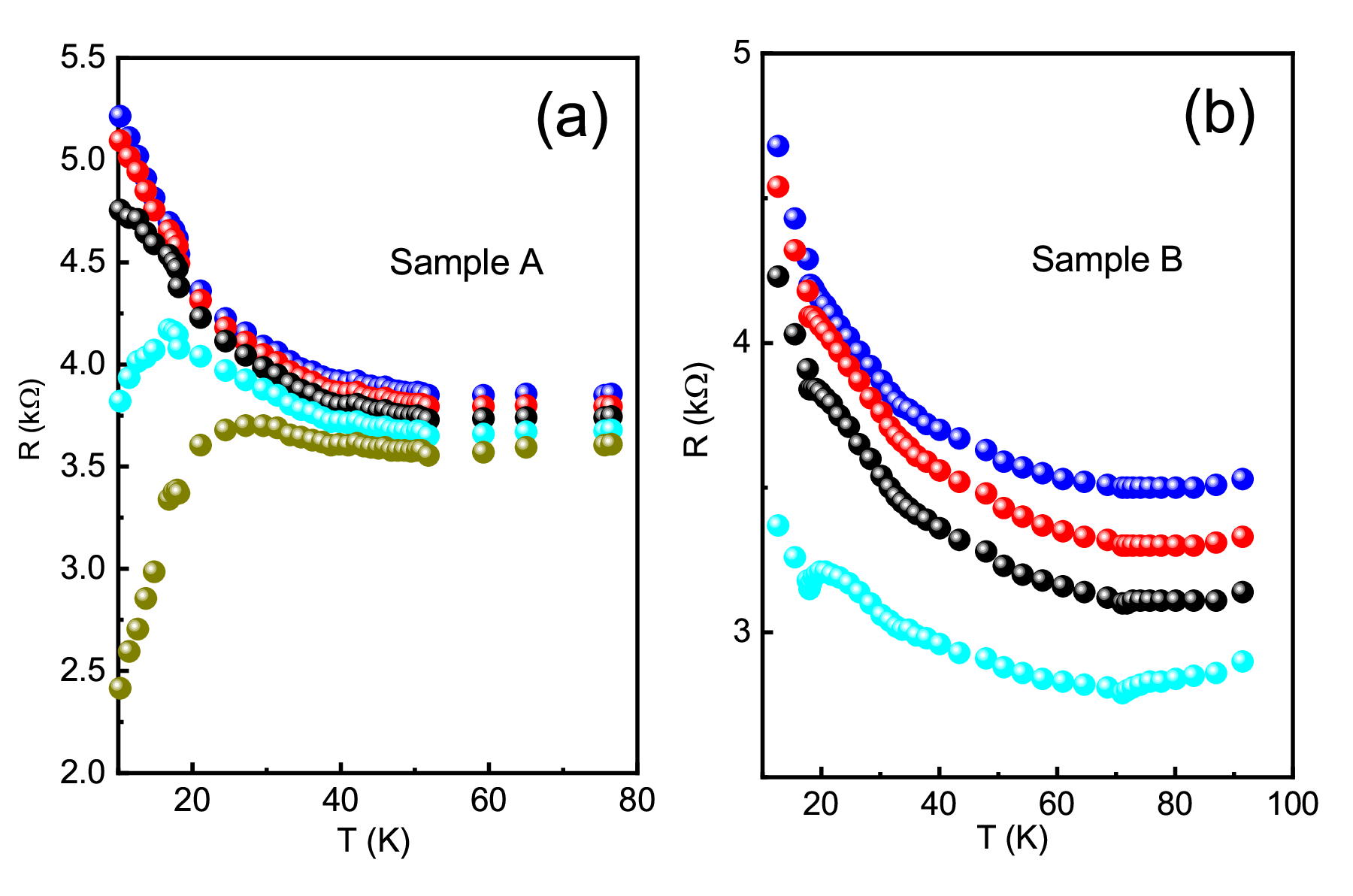}
\caption{(Color online) (a) Resistivity $\rho(T)$ as a function of temperature for different densities for sample A. The total density is
$N_{s}\approx 0$ (charge neutrality point) (blue);  $N_{s}\approx3\times10^{10} cm^{-2}$ (red); $N_{s}\approx6.1\times10^{10} cm^{-2}$ (black); $N_{s}\approx8\times10^{10} cm^{-2}$ (cyan); $N_{s}\approx10\times10^{10} cm^{-2}$ (dark yellow). (b) Resistivity $
\rho(T)$ as a function of temperature for different densities for sample B. The total density is
$N_{s}\approx 0$ (charge neutrality point) (blue);  $N_{s}\approx3\times10^{10} cm^{-2}$ (red); $N_{s}\approx6.1\times10^{10} cm^{-2}$ (black). $N_{s}\approx8\times10^{10} cm^{-2}$ (cyan).}
\end{figure*}
\begin{figure}
\includegraphics[width=9cm]{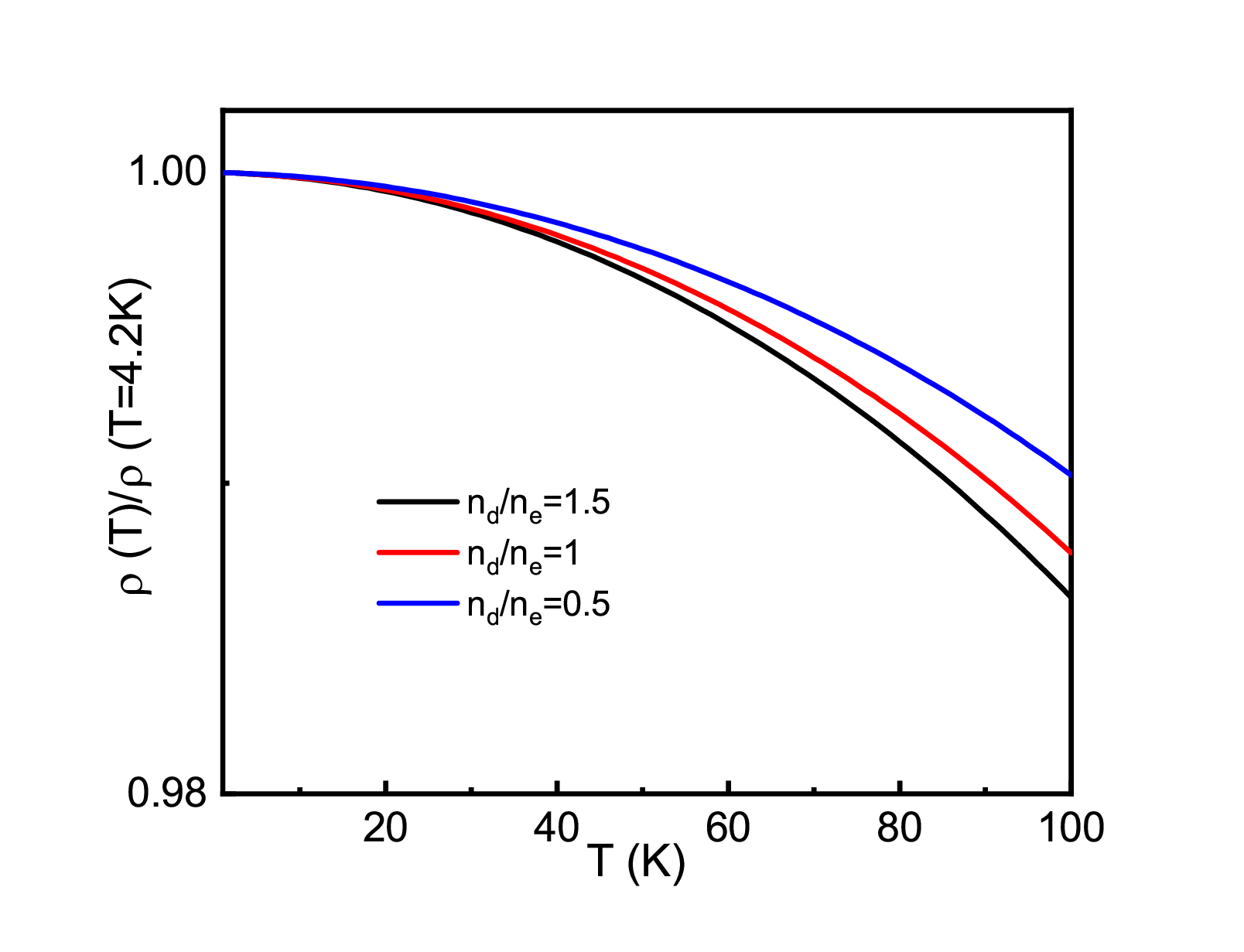}
\caption{(Color online) (a) The resistivity ratio$
\rho(T)/\rho(T=4.2K)$ as a function of temperature for different parameters calculated from Eqs.(5,13,14). Total density $N_{s}=3\times10^{10} cm^{-2}$, the ratio $\tau_{e}/\tau_{d}=5, \tau_{d}=0.4\times10^{-12} s$, the ratio $n_{d}/n_{e} (T=4.2K)=1.5$ (black); 1 (red); 0.5 (blue). Parameter $\beta=1$.}
\end{figure}
Figure 6 depicts the correlation between the relaxation rates $1/\tau_{de}^{}$ and $1/\tau_{ed}^{}$ and temperature, represented by the blue and black lines. The computations employ Equations \ref{eq13} and \ref{eq14} with the parameter $\beta$ set to 1. Due to the integrals $\mathfrak{J_{1,2}}$ behaving approximately like power functions of $T^{\delta}$ with $\delta\approx0.5$, the relaxation rates exhibit distinct temperature dependencies. Specifically, the rate $1/\tau_{de}^{}$ is a weakly temperature-dependent function, while the collision rate $1/\tau_{ed}^{}$ follows a quasi linear $T$ dependence.

The relationship between rates $1/\tau_{de}^{}$ and $1/\tau_{ed}^{}$ depends on the ratio $n_{e}/n_{d}$ and the temperature interval. Notably, over a wide temperature range, the interaction rate maintains the relationship $1/\tau_{de}^{} > 1/\tau_{ed}^{}$ for $n_{e}/n_{d}=1.5$. Moreover, the relaxation rate $1/\tau_{de}^{}$ surpasses $1/\tau_{de}$ at temperatures exceeding 35K for the inverse relation $n_{e}/n_{d}=1.5$. It is essential to note that interactions between massive and massless Dirac electrons conserve the overall momentum density, and the total relaxation rate involved in the transport coefficient is given by $1/\tau_{int}^{*}=1/\tau_{de}^{*}+1/\tau_{ed}^{*}$. The temperature dependence of the relaxation rate $1/\tau_{int}^{*}(T)$ describes an intermediate power with a temperature-dependent exponent index.

In Figure 1c, the chemical potential exhibits a range of variation from 1 to 3 meV for total electron densities in the order of $(1-3)\times 10^{10} cm^{-2}$. Within this density range, we can assert that both massive and massless Dirac electrons adhere to Boltzmann statistics for temperatures between $10 K$ and $100 K$.

Figures 7a and 7b illustrate the temperature dependence of resistance in the vicinity of the charge neutrality point for low electron densities. It is evident that resistance decreases as the temperature increases near the CNP. This trend is attributed to the presence of a small gap in the spectrum, approximately $\Delta \sim 1 meV$, leading to an activation law $R\sim \exp(\Delta/2kT)$ for the resistance, as indicated in \cite{ferreira}. Additionally, this behavior is justified by the quantization of resistance and nonlocal resistance resulting from the helical edge states, confirming the topological insulator nature of the triple HgTe-based quantum well, as demonstrated in our previous publications \cite{ferreira}.  Furthermore, it is crucial to note that although the involvement of helical edge states is significant, especially when the chemical potential resides within the gap, their contribution becomes negligible,  when $\mu$ is shifted to the conductivity band, and the edge states become heavily intertwined with the bulk conductivity. 

Beyond the CNP, we observe weakly temperature-independent resistance, as depicted in Figures 3a and 3b, and more comprehensively in Figures 7a and 7b. Notably, a strong temperature dependence of resistance is discernible far from the CNP, where the resistance increases by more than an order of magnitude with rising temperature. This observation underscores distinct transport regimes in the regions near and far from the CNP—specifically, for low densities where a non-degenerate system can be anticipated, and at high densities where the system becomes degenerate across the entire temperature range.

To assess the agreement of our results with theory in the non-degenerate regime, we computed the resistivity using Equations \ref{eq5} and \ref{eq13}-\ref{eq16}. Figure 8 illustrates these calculations for a total density $N_{s}=3\times10^{10} cm^{-2}$ and various density ratios $n_{d}/n_{e}(T=4.2 K)$. Interestingly, there is a small ($\sim 2\%$) reduction in resistance with increasing temperature. This finding may seem surprising given the distinct growth observed in the scattering time $1/\tau_{int}^{*}(T)$ (see Fig. 6). It is essential to note, however, that in the Boltzmann regime, several parameters are expected to exhibit temperature dependence.

Primarily, the densities of massive and massless electrons strongly depend on temperature in the non-degenerate regime (see \ref{eq12}). Despite this, as previously mentioned, the total density remains fixed by the external gate voltage, suggesting a substantial redistribution of charge carriers as temperature increases. Additionally, the effective mass exhibits strong temperature dependence, significantly contributing to the temperature-dependent resistivity. These temperature-dependent parameters counterbalance the increase in the relaxation rate $1/\tau_{int}^{*}(T)$ and ultimately result in only a weakly temperature-dependent resistance. This effect bears some similarity to the situation in graphene, where the non-degenerate limit has been explored in single-layer and bilayer graphene, with electron-hole pairs being thermally excited \cite{nam, tan}. In these cases, the electron-hole collision rate is expected to be proportional to $T$, leading to temperature-independent conductivity.

It's important to acknowledge that our model, while informative, is overly simplistic for capturing the intricate details observed in the experimental data presented in Figures 7a and 7b, particularly in terms of exact evolution  of the resistance with T. Achieving a more accurate agreement with the experimental results, especially with Equation \ref{eq5}, necessitates a precise understanding of the behavior of the density of states within the gap. 

Additionally, our tight-binding model, though useful, is too rudimentary to accurately characterize parameters in the vicinity of the  CNP, where potential disorder may also contribute to smoothing effects. Consequently, it becomes evident that the weak temperature dependence observed in the experiments near the CNP aligns with our theoretical model to a certain extent. However, it is crucial to recognize the limitations of our model and acknowledge the need for more sophisticated approaches to capture the nuances exhibited by the experimental data.

\section {conclusion}
In summary, our study focused on the temperature-dependent resistivity in a triple HgTe quantum well that accommodates two branches of fermions, one with massive and the other with massless Dirac characteristics. We observed quadratic temperature dependencies resulting from interactions between the Dirac fermions and massive electrons in the fully degenerate regime. In contrast, when  both type of electrons adhered to Boltzmann statistics, resistivity remained weakly temperature-dependent.

Our findings validate that the presented model comprehensively describes the conductivity of the triple quantum well across a broad spectrum of temperatures and carrier densities. In our ultraclean samples, electron-electron scattering takes precedence, exerting a significantly more notable impact than impurity scattering. It also demonstrates the unified nature of hydrodynamic transport across different systems, irrespective of the carriers' sign and spectrum type.

\section{ACKNOWLEDGMENTS}
This work is supported by FAPESP (São Paulo Research  Foundation) Grants No. 2019/16736-2 and No. 2021/12470- 8, CNPq (National Council for Scientific and Technological  Development), and by the Ministry of Science and Higher Education of the Russian Federation and Foundation for the Advancement of Theoretical Physics and Mathematics  “BASIS.”

\end{document}